\documentclass[12pt]{article}
\usepackage[utf8]{inputenc}
\usepackage{graphicx}
\usepackage[labelformat=simple]{subcaption}
\usepackage{rotating}
\usepackage{tikz}
\usepackage{fullpage}
\usepackage{listings}
\usepackage{bm}
\usepackage{comment}
\usepackage{natbib}   

\usepackage{booktabs} 
\usepackage{hyperref}

\usepackage{pdflscape} 
\usepackage{authblk} 
\usepackage[pass]{geometry} 
\usepackage{float} 
\usepackage{enumitem}   
\usepackage{graphicx}

\usepackage[table]{xcolor}
\usepackage{ragged2e}
\newcommand{\covid}{\texttt{COVID-19}}

\newcommand{\covidwave}{\texttt{COVID-19 WAVE }}

\newcommand{\GAMabs}{\texttt{GAM-abs}}
\newcommand{\GAMdiff}{\texttt{GAM-diff}}
\newcommand{\GHQ}{\texttt{GHQ36}}

\makeatletter
\renewcommand{\maketitle}{
  \begin{flushleft}
    {\LARGE \@title \par}
    \vskip 1em
    { \@author}
  \end{flushleft}
}
\makeatother

\makeatletter
\renewcommand{\abstract}{
  \begin{flushleft}
    
  \end{flushleft}
}
\makeatother

\begin{document}

\begin{titlepage}
\title{Modeling mental health trajectories during the COVID-19 pandemic using UK-wide data in the presence of sociodemographic variables}

\author[1,*]{Glenna Nightingale}
\author[2]{Karthik Mohan}
\author[3]{Eloi Ribe}
\author[4]{Valentin Popov}
\author[2]{Shakes Wang}
\author[1]{Clara Calia}
\author[5]{Luciana Brondi}
\author[2]{Sohan Seth}

\affil[1]{School of Health in Social Science, University of Edinburgh, Edinburgh, United Kingdom}
\affil[2]{School of Informatics,University of Edinburgh, Edinburgh, United Kingdom}
\affil[3]{Department of Gerontology, University of Southampton, Southampton, United Kingdom}
\affil[4]{School of Mathematics and Statistics, University of St Andrews, St Andrews, United Kingdom}
\affil[5]{Institute of Global Health and Development, Queen Margaret University, Edinburgh, United Kingdom}

\affil[ *]{Corresponding author name and email address: Glenna Nightingale - \url{Glenna.Nightingale@ed.ac.uk}}

\maketitle



\begin{abstract}
\textbf{Abstract:} 
\textbf{Background:} The negative effects of the {\covid} pandemic on the mental health and well-being of populations are an important public health issue. Although several studies have reported on these effects in the UK, less is known about temporal trends and how different demographic groups were affected. Our study aims to determine the underlying factors shaping mental health trajectories during the {\covid} pandemic in the UK.

 \textbf{Methods:} Data from the Understanding Society {\covid} Study, from April $2020$ to September $2021$ were utilized. The core analysis included $17,961$ individuals (aged $16$ and over)  with a total of $179,610$ observations, at $9$ time points, focusing on the General Health Questionnaire ({\GHQ}) scores for mental health outcomes. We used Generalized Additive Models to evaluate trends over time and the role of sociodemographic variables, i.e., age, sex, ethnicity, country of residence (in UK), job status (employment), household income, living with a partner, living with children under age 16, and living with a long-term illness, on the variation of mental health during the study period.

\textbf{Results:}  Statistically significant differences in mental health were observed for age, sex, ethnicity, country of residence (in UK), job status (employment), household income, living with a partner, living with children under age 16, and living with a long-term illness. To summarize some of these, women experienced higher {\GHQ} scores relative to men with the {\GHQ} score expected to increase by $1.260\ (95\% \textrm{CI:}\,  1.176,1.345)$. Individuals living without a partner were expected to have higher {\GHQ} scores, of $1.050\ (95\% \textrm{CI:}\,  0.949, 1.148)$ more than those living with a partner, and age groups 16-34, 35-44, 45-54, 55-64 experienced higher {\GHQ} scores relative to those who were 65+.  Finally, individuals, with relatively lower household income were likely to have poorer mental health relative to those who were more well off. 

 \textbf{Conclusion:} This study identifies key demographic determinants shaping mental health trajectories during the {\covid} pandemic in the UK. Policies aiming to reduce mental health inequalities should target women, youth, individuals living without a partner, individuals living with children under 16, individuals with a long-term illness, and lower income families.

 \textbf{keywords:} mental health COVID-19, UKHLS
\end{abstract}

\end{titlepage}

\section{Introduction}
The impact of coronavirus disease ({\covid}) on mental health is undeniably varied between different demographic groups \citep{rahman2021suicidal}  and the adverse mental health consequences of the {\covid} pandemic have been widely documented in recent years. However, the complete extent of the psychosocial impact of the {\covid} pandemic across the globe is still unknown and its long-term impact is a topic of active research \citep{daly2022longitudinal}. Studies have suggested that the {\covid} pandemic has resulted in increased activation of the Behavioral Immune System \citep{makhanova2020behavioral} and the Stress Response System \citep{taylor2006hypochondriasis}, highlighting the risk that this pandemic poses to mental health. The wider societal impact of the pandemic can also be understood using the ADAPT model \citep{silove2013adapt}, which proposes five core psychosocial pillars as critical elements of any healthy society, and the {\covid} pandemic tends to substantially affect the five psychosocial pillars of the ADAPT model, namely safety and security, roles and identities, bonds and networks, existential meaning and justice. Therefore, it is not surprising that several studies have reported that the impact of {\covid} on mental health varies significantly according to different demographic characteristics (sex, age, etc.) of the population studied \citep{coco2023psychosocial}.

{\covid} can adversely affect mental health directly through neuropsychiatric sequelae (both during acute disease and as a long-term effect) or indirectly as a consequence of strict public health control measures (e.g. lockdowns) that lead to social disruption \citep{mollica2004mental}. In the UK, evidence suggests that both short-term and long-term mental health harms related to the {\covid} pandemic are substantial, complex and concerning \citep{wels2022mental}. Aside from the documented negative impacts, some studies suggest that certain individuals, such as middle-aged adults in secure jobs who went on furlough during the pandemic, experienced positive mental health \citep{perelli2020couples, perelli2023better}. Other studies have suggested that individuals, who were able to spend time with immediate family whilst being at very low risk of serious illness from {\covid}, experienced improved mental health.
 \cite{wang2024impact} indicates that the mental health of men was at a higher advantage in terms of being protected during furlough.   In particular, \cite{perelli2023better} suggested that the UK furlough scheme contributed to the protection of couples’ relationship during the pandemic. \cite{pierce2021mental} indicated that the onset of the {\covid} pandemic initiated a wave of decline in population mental health in the UK, and that this decline was unequal amongst various demographic groupings including sex and age. 

Several studies have used the Understanding Society {\covid} Study data, which is based on the UK Household Longitudinal Study (UKHLS), a UK-wide longitudinal survey. For example, ethnic differences in mental health were reported by \citep{rahman2021suicidal}. Other studies \citep{davillas2021first} using the UKHLS data, finds a decline in mean population mental health in the UK at the onset of the pandemic from 24-30th April 2020. \citep{pierce2021mental} have used Latent Class Mixture Models to report five distinct mental health trajectories during the pandemic using the first six waves from the UKHLS {\covid} study.  Following these studies, we use more UKHLS {\covid} waves, i.e., nine, to explore the trend of mental health during a longer period between April 2020 to September 2021, and the effect of different sociodemographic variables, particularly, age (banded), sex, ethnicity (white or non-white), country of residence, 
job status (employment), 
household income (income quintile),
living with a partner, 
living with children under 16, and 
having a long-term illness, on mental health with the goal of providing evidence to public health policymaking.

\section{Data}\label{data}

 Our study uses nine Understanding Society {\covid} Study survey waves covering 18 months from April $2020$ to September $2021$. The nine waves correspond to April $2020$ (\covidwave 1), May $2020$ (\covidwave 2), June $2020$ (\covidwave 3), July $2020$ (\covidwave 4), September $2020$ (\covidwave 5), November $2020$ (\covidwave 6), January $2021$ (\covidwave 7), March $2021$ (\covidwave 8), and September $2021$ (\covidwave 9), respectively. We also use a baseline period, from $2018$ to $2019$ (Wave 0) for our analysis.  This baseline was obtained from the main panel data, and is not available in the {\covid} waves. For ease of reference, we refer to the \covidwave 1-9 as simply ‘waves 1-9’, and we refer to the baseline period as ‘baseline wave’ (or `wave 0').  Variables used in the study are indicated in Table \ref{variables}.  All individuals aged 16 and above in April $2020$ are included in the analysis.

 We focus on the General Health Questionnaire ({\GHQ}) \citep{jackson2017hair} as outcome variable which is on a 36-point scale. {\GHQ} is measure of non-psychotic psychiatric cases of mental health of adult individuals in the population. The 36-point scale denoted by {\GHQ} is based on 12 questions each with (ordinal) responses which range from 1-4 and in the Understanding Society {\covid} Study, these responses are recorded on a scale of 0-3 instead. The series of twelve questions screen for minor psychiatric disorders in the past few weeks, and a higher {\GHQ} score denotes poorer mental health \citep{jackson2017hair}. This measure is widely used and has been validated in general and clinical populations. It has also been used in previous analysis of the impact of {\covid} on mental health  \citep{davillas}. Table \ref{tab:char_apr_nov} represents the characteristics of the participants by wave.

\begin{sidewaystable}[]
\tiny
\caption{Participant characteristics for covid waves 1-6 representing Apr 2020, May 2020, June 2020, July 2020, Sept 2020, and Nov 2020 respectively}
\label{tab:char_apr_nov}
\begin{tabular}{|l|l|l|l|l|l|l|l|}

\hline
\textbf{Variable} &\textbf{baseline} & \textbf{Apr 20} & \textbf{May 20} & \textbf{Jun 20} & \textbf{Jul 20} & \textbf{Sep 20} & \textbf{Nov 20} \\


\hline
\textbf{Sex} & & & & & & & \\
Male & 7,584 (42\%) & 6,862 (42\%) & 5,768 (42\%) & 5,484 (41\%) & 5,394 (42\%) & 5,076 (42\%) & 4,756 (42\%) \\
Female & 10,377 (58\%) & 9,407 (58\%) & 8,074 (58\%) & 7,733 (59\%) & 7,491 (58\%) & 7,050 (58\%) & 6,605 (58\%) \\
Unknown & 0 & 1,692 & 4,119 & 4,744 & 5,076 & 5,835 & 6,600 \\
\hline



\textbf{Ageband} & & & & & & & \\ 
16--34 & 3,635 (20.2 \%) & 2,918 (18.2\%) & 2,066 (14.9\%) & 1,873 (14.2\%) & 1,796 (13.9\%) & 1,539 (12.7\%) & 1,375 (12.1\%) \\

35--44 & 2,994 (16.7\%) & 2,574 (16.0\%) & 2,049 (15.0\%) & 1,910 (14.0\%) & 1,855 (14.0\%) & 1,664 (14.0\%) & 1,517 (13.0\%) \\

45--54  &3,725 (20.7\%) & 3,313 (20.0\%) & 2,811 (20.0\%) & 2,623 (20.0\%) & 2,545 (20.0\%) & 2,385 (20.0\%) & 2,193 (19.0\%) \\

55--64 & 3,645 (20.3\%) & 3,372 (21.0\%) & 3,040 (22.0\%) & 2,988 (23.0\%) & 2,914 (23.0\%) & 2,819 (23.0\%) & 2,683 (24.0\%) \\

65+ & 3,961 (22\%)& 4,106 (25.5\%) & 3,876 (28.3\%) & 3,835 (29.0\%) & 3,784 (28.8\%) & 3,729 (30.4\%) & 3,602 (31.6\%) \\

Unknown & 1 (0.0\%) & 1,678 & 4,119 & 4,732 & 5,067 & 5,825 & 6,591 \\

\hline

\textbf{Ethnicity (White / Non-White)} & & & & & & & \\
0 & 15,483 (87\%) & 15,483 (87\%) & 15,483 (87\%) & 15,483 (87\%) & 15,483 (87\%) & 15,483 (87\%) & 15,483 (87\%) \\
1 & 2,344 (13\%) & 2,344 (13\%) & 2,344 (13\%) & 2,344 (13\%) & 2,344 (13\%) & 2,344 (13\%) & 2,344 (13\%) \\
Unknown & 134 & 134 & 134 & 134 & 134 & 134 & 134 \\
\hline

\textbf{Country} & & & & & & & \\
England & 14,531 (81\%) & 0 (NA\%) & 11,190 (81\%) & 10,673 (81\%) & 10,397 (81\%) & 9,809 (81\%) & 9,225 (81\%) \\
Wales & 1,067 (5.9\%) & 0 (NA\%) & 867 (6.3\%) & 829 (6.3\%) & 811 (6.3\%) & 742 (6.1\%) & 687 (6.0\%) \\
Scotland & 1,559 (8.7\%) & 0 (NA\%) & 1,214 (8.8\%) & 1,173 (8.9\%) & 1,129 (8.8\%) & 1,076 (8.9\%) & 994 (8.7\%) \\
N.Ireland & 795 (4.4\%) & 0 (NA\%) & 571 (4.1\%) & 554 (4.2\%) & 557 (4.3\%) & 509 (4.2\%) & 464 (4.1\%) \\
Unknown &9 & 17,961 & 4,119 & 4,732 & 5,067 & 5,825 & 6,591 \\

\hline

\textbf{Job status (Employment)} & & & & & & & \\
Yes & 11,142 (62\%) & 9,502 (61\%) & 8,226 (60\%) & 7,729 (59\%) & 7,470 (58\%) & 6,887 (57\%) & 6,393 (57\%) \\
No & 6,735 (38\%) & 6,039 (39\%) & 5,526 (40\%) & 5,401 (41\%) & 5,359 (42\%) & 5,198 (43\%) & 4,901 (43\%) \\
Unknown & 84 & 2,420 & 4,209 & 4,831 & 5,132 & 5,876 & 6,667 \\

\hline

\textbf{Household income (Quintiles)} & & & & & & \\
Q1 (lowest) & 851(4.7\%) & NA & 2,304 (20.2\%) & 2,101 (19.4\%) & 2,741 (26.5\%) & 2,719 (27.7\%) & 2,309 (26.4\%) \\
Q2 & 3693 (20.6\%) & NA & 2,570 (22.5\%) & 2,409 (22.2\%) & 2,055 (19.9\%) & 1,853 (18.9\%) & 1,663 (19.0\%) \\
Q3 & 4708 (26.2\%) & NA & 2,356 (20.6\%) & 2,223 (20.5\%) & 1,924 (18.6\%) & 1,771 (18.0\%) & 1,479 (16.9\%) \\
Q4 & 5038 (28\%) & NA & 2,223 (19.4\%) & 2,130 (19.6\%) & 1,829 (17.7\%) & 1,640 (16.7\%) & 1,463 (16.7\%) \\
Q5 (highest) & 3668 (20.4\%) & NA & 1,978 (17.3\%) & 1,994 (18.4\%) & 1,797 (17.4\%) & 1,844 (18.8\%) & 1,840 (21.0\%) \\
Unknown & 3 & 17,961 & 6,530 & 7,104 & 7,615 & 8,134 & 9,207 \\

\hline

\textbf{Availability of job flexibility} & & & & & & & \\
No & 7,648 (81\%) & 0 (NA\%) & 0 (NA\%) & 4,945 (75\%) & 0 (NA\%) & 4,306 (73\%) & 0 (NA\%) \\
Yes & 1,842 (19\%) & 0 (NA\%) & 0 (NA\%) & 1,656 (25\%) & 0 (NA\%) & 1,565 (27\%) & 0 (NA\%) \\
Unknown & 8,471 & 17,961 & 17,961 & 11,360 & 17,961 & 12,090 & 17,961 \\

\hline

\textbf{Use of job flexibility} & & & & & & & \\
No & 6,811 (88\%) & 0 (NA\%) & 0 (NA\%) & 3,867 (80\%) & 0 (NA\%) & 3,288 (77\%) & 0 (NA\%) \\
Yes & 904 (12\%) & 0 (NA\%) & 0 (NA\%) & 986 (20\%) & 0 (NA\%) & 963 (23\%) & 0 (NA\%) \\
Unknown & 10,246 & 17,961 & 17,961 & 13,108 & 17,961 & 13,710 & 17,961 \\
\hline

\hline
\textbf{On furlough} & & & & & & & \\
No & 0 (NA\%) & 6,664 (80\%) & 5,559 (94\%) & 5,146 (97\%) & 5,007 (98\%) & 0 (NA\%) & 0 (NA\%) \\
Yes & 0 (NA\%) & 1,615 (20\%) & 371 (6.3\%) & 175 (3.3\%) & 80 (1.6\%) & 0 (NA\%) & 0 (NA\%) \\
Unknown & 17,961 & 9,682 & 12,031 & 12,640 & 12,874 & 17,961 & 17,961 \\

\hline

\textbf{Living with a partner} & & & &  & & & \\
Yes & 5,558 (31\%) & 11,712 (72\%) & 9,753 (70\%) & 9,367 (71\%) & 9,124 (71\%) & 8,661 (71\%) & 8,056 (71\%) \\
No  & 12,403 (69\%) & 4,571 (28\%) & 4,089 (30\%) & 3,862 (29\%) & 3,770 (29\%) & 3,475 (29\%) & 3,313 ((29\%))\\
Unknown & 0 & 1,678 & 4,119 & 4,732 & 5,067 & 5,825 & 6,592	\\
\hline

\textbf{Children under 16} & & & & & & & \\
None & 13,317 (74\%) & 14,690 (90\%) & 12,731 (92\%) & 12,215 (92\%) & 11,920 (92\%) & 11,259 (93\%) & 10,579 (93\%) \\
1 child & 1,903 (11\%) & 1,239 (7.6\%) & 874 (6.3\%) & 810 (6.1\%) & 761 (5.9\%) & 690 (5.7\%) & 623 (5.5\%) \\
2 or more & 2,681 (15\%) & 354 (2.2\%) & 237 (1.7\%) & 204 (1.5\%) & 213 (1.7\%) & 187 (1.5\%) & 167 (1.5\%) \\
Unknown & 60 & 1,678 & 4,119 & 4,732 & 5,067 & 5,825 & 6,592 \\
\hline

\textbf{Long term illness} & & & & & & & \\
No & 11,836 (66\%) & 8,159 (50\%) & 7,358 (53\%) & 6,469 (49\%) & 6,194 (48\%) & 5,498 (45\%) & 5,035 (44\%) \\
Yes & 6,089 (34\%) & 8,124 (50\%) & 6,484 (47\%) & 6,760 (51\%) & 6,700 (52\%) & 6,638 (55\%) & 6,335 (56\%) \\
Unknown & 36 & 1,678 & 4,119 & 4,732 & 5,067 & 5,825 & 6,591 \\
\hline

\textbf{Being pregnant} & & & & & & & \\
No & 4,757 (98\%) & 4,129 (97\%) & 3,285 (100\%) & 3,046 (100\%) & 2,911 (99\%) & 2,593 (99\%) & 2,373 (99\%) \\
Yes & 100 (2.1\%) & 106 (2.5\%) & 13 (0.4\%) & 14 (0.5\%) & 16 (0.5\%) & 16 (0.6\%) & 19 (0.8\%) \\
Unknown & 13,104 & 13,726 & 14,663 & 14,901 & 15,034 & 15,352 & 15,569 \\
\hline

\textbf{GHQ} & 11.0 (8.0, 15.0) & 11.0 (8.0, 15.0) & 11.0 (8.0, 14.0) & 11.0 (8.0, 13.0) & 11.0 (8.0, 13.0) & 12.0 (9.0, 15.0) \\
Unknown & 3,090 & 4,438 & 5,074 & 5,371 & 6,232 & 6,909 \\ 
\hline 

\end{tabular}

\end{sidewaystable}

We have considered thirteen variables namely: sex, age (banded), ethnicity (white or non-white), country of residence,
job status (employment), 
household income (income quintile),
availability (and separately, the use of) job flexibility, 
being on furlough, 
living with a partner, living with children under 16 years of age, 
having long-term illness, and being pregnant; among which nine with sufficient non-missingness were used in the final model (see Additional files 3 -5 for Missingness summary for Sex, Ageband, Ethnicity, and Country across COVID-19, Missingness summary for Job status (Employment), Household income (Quintiles),
Availability of job flexibility, and Use of job flexibility across COVID-19 waves, and Missingness summary for variables: On furlough, Living with a partner, Children under 16, Long-term illness, Being pregnant, and GHQ-36. Waves a–j represent April 2020 to September 2021
).  The variables excluded in both models exhibited missingness percentages over $50\%$.

\begin{sidewaystable}[]
\small
\caption{Variable Characteristics. Static variables do not change over waves while dynamic variables do as well as outcome.}
\begin{tabular}{|l|l|l|l|}\hline
Variables                     & Description                                           & Values   Range                  & In models \\ \hline
\textbf{Static}                        &                                                       &                                 &                    \\\hline
Sex                           & Respondent’s sex                                      & Male/Female                     & Yes                \\\hline
Ethnicity                     & Binary grouping of ethnicity; white/non-white         & White/Non-white                 & Yes                \\\hline
\textbf{Dynamic}                      &                                                       &                                 &                    \\\hline
Age band                      & Age grouped into 5 bands for modeling                 & 16–34, 35–44, 45–54, 55–64, 65+ & Yes                \\\hline
Country                       & Country of residence (UK)                             & Eng, Wales, Scot, N Ireland     & Yes                \\\hline
Job status                    & Employment status                                     & Employed/Unemployed             & Yes                \\\hline
Income quintile               & Household income grouped into quintiles               & Q1, Q2, Q3, Q4, Q5              & Yes                \\\hline
Living with a partner         & Whether living with a partner                         & Yes/No                          & Yes                \\\hline
Living with children under 16 & Number of children under 16 in the household          & None, 1, 2 or more              & Yes                \\\hline
Long-term illness             & Presence of a long-term health condition              & Yes/No                          & Yes                \\\hline
Job flexibility               & Whether job flexibility is available                  & Yes/No                          & No                 \\\hline
Use of Job Flexibility        & Whether job flexibility was used                      & Yes/No                          & No                 \\\hline
Whether pregnant              & Whether pregnant                                      & Yes, No, Don’t know             & No                 \\\hline
Whether on furlough           & Furloughed under the Coronavirus Job Retention Scheme & Not furloughed, Furloughed      & No                 \\\hline
\multicolumn{2}{l}{\textbf{Outcome} }                                                          &                                 &                    \\\hline
GHQ                           & General Health Questionnaire (GHQ36).                 & 0 to 36                         & Yes     \\\hline          
\end{tabular}

\label{variables}
\end{sidewaystable}

\newpage
\section{Methods}\label{sec:GAMs}

We fitted the following models:  (i) a Generalized Additive Model to predict {\GHQ} over all nine waves simultaneously, referred to as {\GAMabs}, and (ii) a Generalized Additive model to predict the change in {\GHQ} over all nine waves from the baseline wave simultaneously, referred to as {\GAMdiff}.
First, in {\GAMabs}, we focus on modelling the {\GHQ} score directly as this allows us to study the dynamics of the general trend and the effect of the demographic variables on mental health during the pandemic.  Second, in {\GAMdiff}, we focus on modelling the change in {\GHQ} score from the baseline, as this allows understanding the demographics that were affected during the pandemic compared to their pre-pandemic state. 

We construct the {\GAMabs} as:
\begin{equation} \label{GAMabs2}
\GHQ_{it} = \beta_{0} + s(t) + \bm{\beta}^\top \bm{x}_{it} + \epsilon_{it}
\end{equation}
where $\bm{x}$ denotes the selected nine variables,
$\epsilon_{it}$ denotes noise, assumed to be normally distributed, and we model the nonlinear variation over waves, i.e., $s(t)$ using B-splines of order 3 and with 3 equispaced internal knots (i.e., 6 coefficients, 7 including $\beta_0$). 
The {\GAMdiff} model is built in a similar way; however, the response variable is now the difference of the current {\GHQ} score and the pre-stress baseline level for each individual, that is,
\begin{equation} \label{GAMdiff2}
\GHQ_{it} - \GHQ_{it}^{baseline}= \beta_{0} + s(t) + \bm{\beta}^\top\bm{x}_{it} + \epsilon_{it}
\end{equation}

 \section{Results}
 
Out of 13 variables considered,  4 variables were not included in the analyses since they were available for less than 5 waves. These variables are, ``on furlough", ``being pregnant", ``availability of job flexibility" and ``use of job flexibility".  For verage missingness percentage for variables excluded from GAM models across
COVID-19 waves see Additional file 6. The analysis ready dataset included $13,552$ individuals (aged $16$ and over)  with a total of $69,095$ observations, at $9$ time points for the {\GAMabs} model and $15,708 $ individuals (aged $16$ and over) with a total of $94,017$ observations, at $9$ time points for the {\GAMdiff} model. The relative proportion of different demographics varies across the waves, but in general, the dataset had more females than males, more older people (65+) than younger people (16-34), more whites than non-whites, more individuals who lives in England than the other countries, more employed individual than unemployed, more individuals who are living with partner than not, more individual not living with children under 16 than are, and almost equal number of individuals living with or without long term illness.



\begin{figure}[htbp]
\centering
  \includegraphics[width= 1 \linewidth]{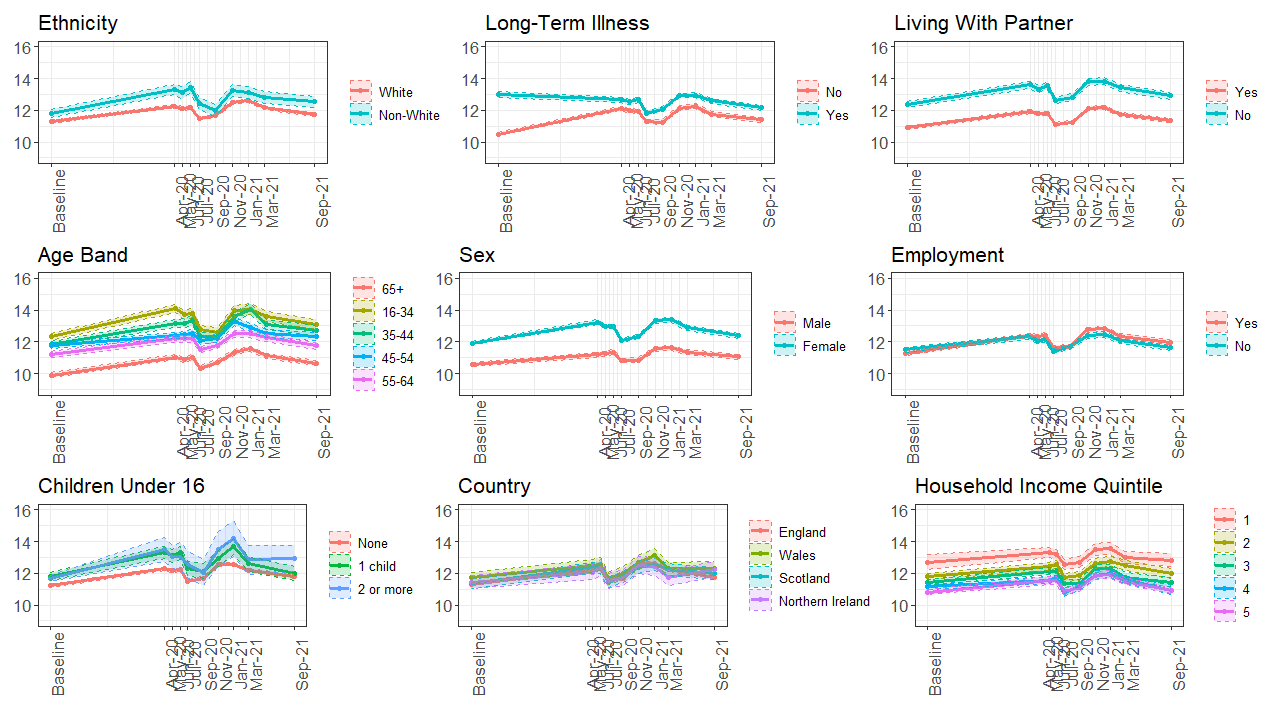} 
\caption{Trend plots for all nine variables in GAMabs model}
\label{trend:figs}
\end{figure}

\begin{figure}[htbp]
\centering
  \includegraphics[width= 1 \linewidth]{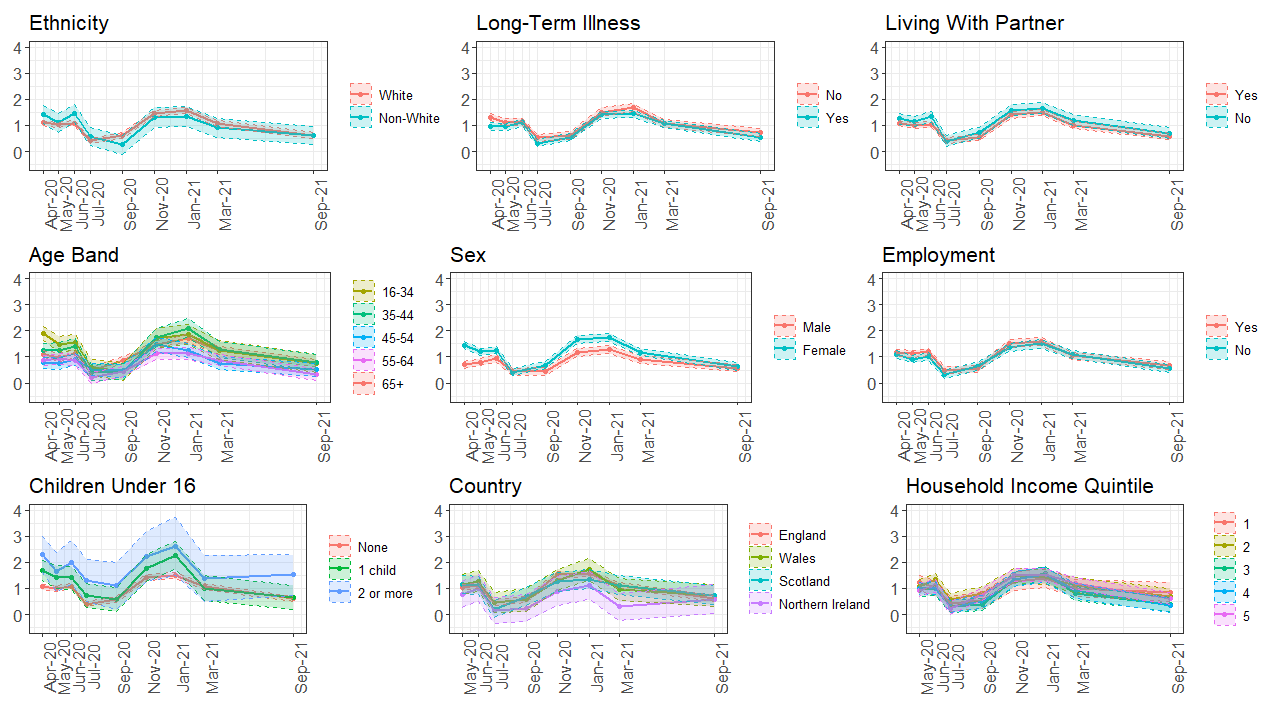}
\caption{Trend plots around baseline value  accompanied by $95\%$ confidence intervals for variables studied.}
  \label{images2}
\end{figure}


Figure \ref{trend:figs}  depicts the trends observed the nine variables incorporated in the analyses, and Figure \ref{images2} illustrates the trend in the difference in mental health scores over the study timeframe for these nine variables. On visual inspection, on average, mental health appears to deteriorate during the winter months and improve in the summer months, and mental health at the beginning of the pandemic was as bad as in the winter months. Second, for each demographic variable (e.g., age banded) the different groups (i.e., 16-24, etc.) show consistent effect over different waves (e.g., 16-34 age group has worse mental health in each wave). Third, the relative effect of the strata on mental health observed in the pre-pandemic baseline wave persists in the {\covid} waves, and finally, the mental health in each pandemic wave on average is worse than their respective pre-pandemic baseline for each groups, i.e., the difference in {\GHQ} score from baseline is positive.

\begin{figure}[t!]
\centering
\includegraphics[width=1\textwidth]{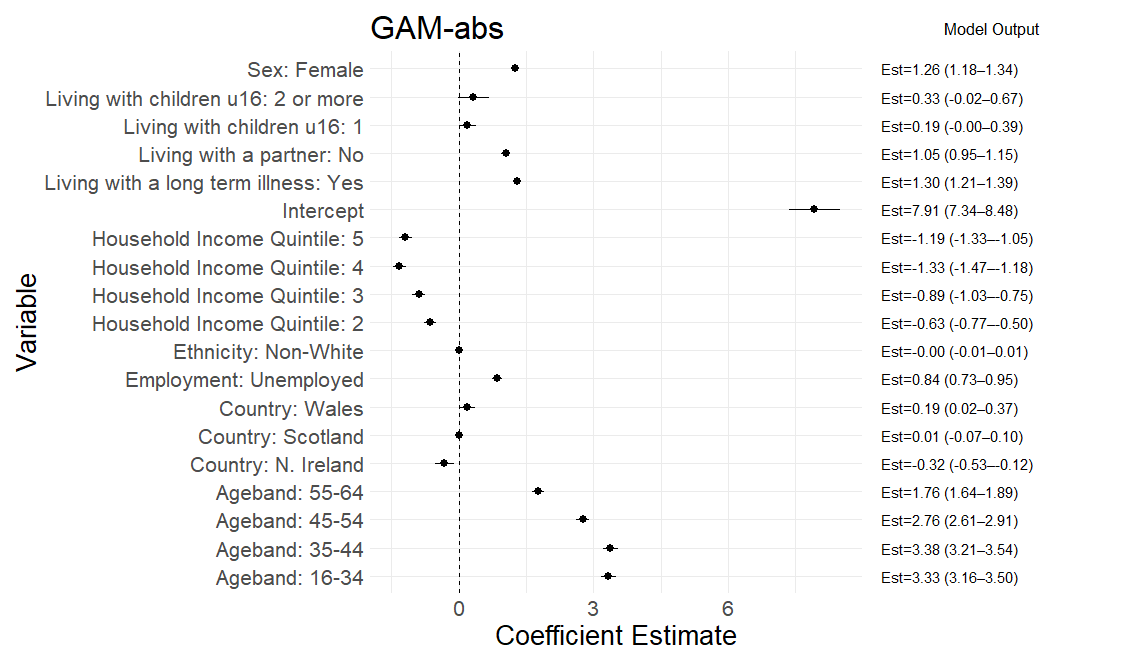}
(a) Forest plots for the \GAMabs model
\includegraphics[width=1\textwidth]{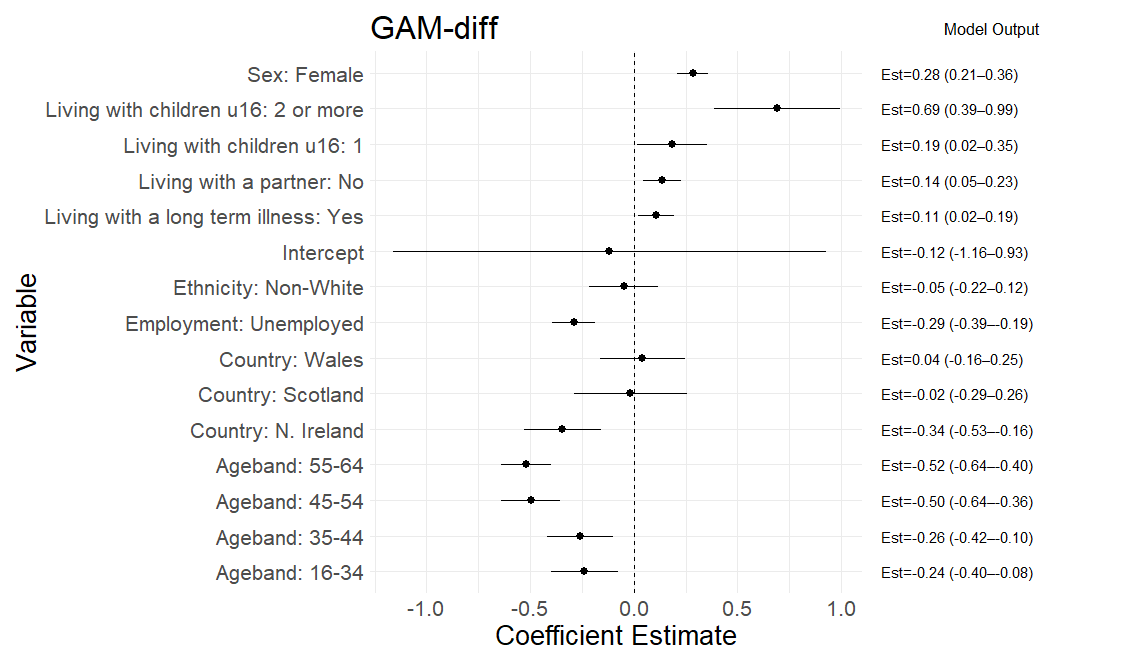}
(b) Forest plots for the \GAMdiff model
\caption{Forest plots for the GAM models}
\label{GAMs}
\end{figure}

\begin{figure}[htbp]
\centering
\includegraphics[width=0.5\textwidth]{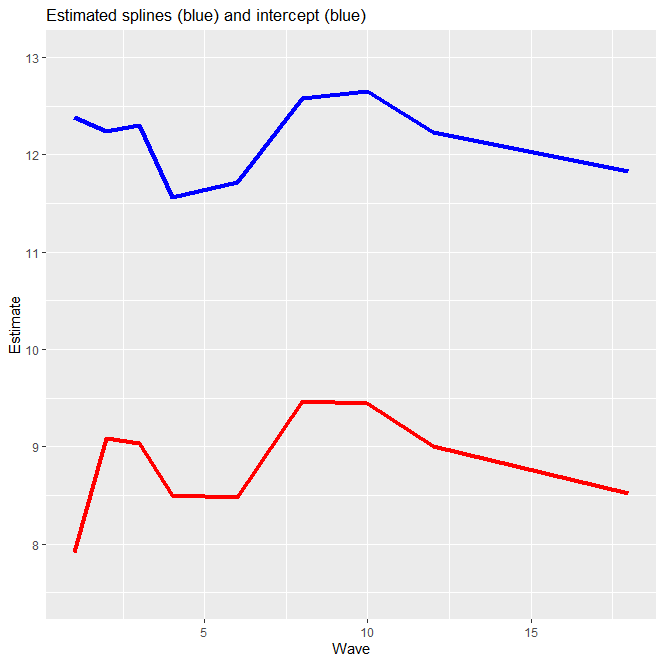}
\caption{Plots of splines from the \GAMabs \ model}
\label{A4}
\end{figure}

\begin{figure}[htbp]
\centering
\includegraphics[width=0.5\textwidth]{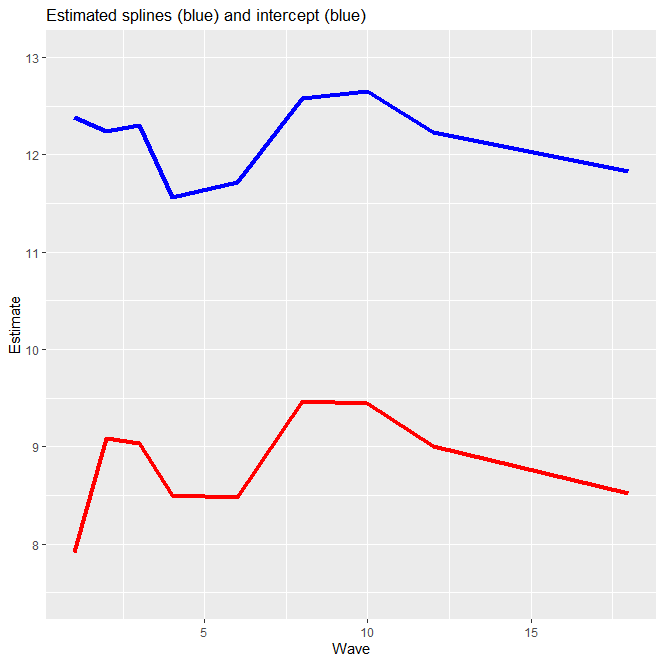}
\caption{Plots of splines from the \GAMdiff \ model}
\label{A5}
\end{figure}

The forest plots for the GAM models (n=14084 for \GAMabs and n=13869 for \GAMdiff) are shown in Figure \ref{GAMs}. The spline plots are provided in Figures \ref{A4} and \ref{A5}.

\paragraph{\GAMabs:} The effects of 7 of the 9 covariates in the model are significant [See Additional file 1 for Results from the \GAMabs \ model. 
\begin{enumerate}[label=(\roman*)]

\item women on an average have $1.26\ (95\% \textrm{CI:}\,   1.176, 1.345)$ points higher {\GHQ} compared to men,
\item Every other age group on average have higher {\GHQ} score compared to $65+$ with 16-34 age group having $3.33\ (95\% \textrm{CI:}\,   3.158, 3.507
)$ points more, $35-44$ age group having $3.38\ (95\% \textrm{CI:}\,   3.210, 3.545
)$ points more than the 65+ age group. 
\item Individuals not living with a partner on an average have $1.05\ (95\% \textrm{CI:}\,   0.949, 1.148)$ points higher {\GHQ} compared to individuals living with partner.
\item For parents with children under 16 we did not find statistically significant associations.
\item Unemployed people on average have $0.84\ (95\% \textrm{CI:}\,  0.730,0.955)$ points higher {\GHQ} compared to employed people.
\item People with long-term disability or illness were found to score $1.30\ (95\% \textrm{CI:}\,  1.210, 1.387
)$ points higher in {\GHQ} compared to those who did not report a long-term illness.
\item When considering income quintiles, we note that the quintiles (Q2 to Q5) were associated with lower {\GHQ} scores relative to Q1, the lowest income quintile.  The {\GHQ} score dropped almost monotonically with increasing quintiles.
\item We note that people in Wales have {\GHQ} scores which are $0.19\ (95\% \textrm{CI:}\,  0.015,0.366)$ points higher than those observed in England.  In contrast, individuals in Northern Ireland have {\GHQ} scores that are $1.33\ (95\% \textrm{CI:}\,  -0.53,-0.119)
$ points less than that observed for the reference category, England.
\item Participants who were of a non white ethnicity experienced a relatively lower {\GHQ} score, but this is only marginally statistically significant ($p=0.048$).

\end{enumerate}

\paragraph{\GAMdiff:}
 The effects of 6 of the 8 covariates in the model are significant. Note that unlike in the previous GAM model, {\GAMabs}, we have not included the household income (income quintile) variable due to convergence issues.  [See Additional file 2 for Results from the \GAMdiff \ model]
 
\begin{enumerate}[label=(\roman*)]
\item We notice a monotonic decrease in the {\GHQ} difference (relative to age $65+$) as the age group is progressively older (from 16-34 through to 55-64).  

\item An increase in {\GHQ} difference of $0.284\ (95\% \textrm{CI:}\,  0.209, 0.359)$ was observed for women relative to men. 
\item We did not observe any statistically significant associations within the ethnicity variable denoting whether the participant was white or non-white.
\item With England as reference, there were not statistically significant associations for Wales and Scotland.  For Northern Ireland, a relatively lower change, $-0.344\ (95\% \textrm{CI:}\,  -0.528, -0.159)$,  in {\GHQ} scores were observed.
\item Unemployed individuals on an average have a smaller change in {\GHQ} of $-0.2897\ (95\% \textrm{CI:}\,  -0.386, -0.193)$ points compared to employed. 
\item Individuals not living with a partner on average have a higher change in {\GHQ} of $0.136\ (95\% \textrm{CI:}\,  0.052, 0.22)$ points compared to individuals living with a partner.   
\item Parents with children under 16 experienced a relatively greater change in {\GHQ} scores than parents without children under 16.  In particular, an increase in {\GHQ} difference of $0.1857\ (95\% \textrm{CI:}\,  0.02, 0.352)$ was observed for parents with one child under 16 compared to parents with none.  Additionally, a greater increase (0.501) in {\GHQ} difference was observed (than that for parents with one child) for parents with two or more children.
\item Individuals with a long-term illness or disability experienced a greater change in {\GHQ} of $0.105\ (95\% \textrm{CI:}\,  0.026, 0.185)$ points relative to those who did not have a long-term illness.  .
\end{enumerate}

 \section{Discussion}

 Our study has investigated the trends in mental health in the United Kingdom during the {\covid} pandemic and has identified some trends that are valuable for public health policy. First and foremost, our results show that there was a nonlinear pattern in mean mental health scores across the 9 study waves with the worse mental health being observed on average in March 2021 and between November 2020 and January 2021. Finally, there is a noticeable trough in the mean {\GHQ} scores in July 2020 that also coincides with the relatively lower national {\covid} rates, the relaxation of lockdown measures, and with summer. Our two models explore two different aspects of the influence of sociodemographic variables on the {\GHQ} scores: while {\GAMabs} explores which groups (e.g., male or female) had worse mental health during the 9 waves, {\GAMdiff} explores which groups were more affected during this period compared to their baseline mental health, provided the mental health of all groups (e.g., male and female) worsensed during this period as we have observed in our exploratory analysis. 

 \subsection{Age}
Younger adults (age 16-34 in particular) experienced worse mental health (relative to older adults) during the pandemic, while older adults (65+) experienced worsened mental health in relation to their pre-pandemic baseline. 
Various researchers and practitioner groups have suggested that the psychosocial development of younger adults was disrupted by the pandemic restrictions \citep{smyth2022disrupted}.  In particular, adolescence is a critical cognitive and social life stage and the psycho-social implications of the pandemic caused obstacles to positive development.
A global study conducted by the International Labor \citep{barford2021youth} found that youth whose rights were impacted during {\covid}, such as the right to housing or information, expressed worse mental well-being and increased levels of depression and anxiety. Additionally, it has been reported \citep{colombo2024effect} that for middle aged adults who possessed cognitive reserve (CR), their mental health was “protected” during the {\covid} pandemic and that overall, CR acts as a protection during the aging process.




While younger adults consistently reported poorer mental health overall, it is notable that their relative changes (which was negative -- a reduction in {\GHQ}) from baseline were smaller compared to older adults. This suggests that although their mental health was already at risk pre-pandemic, their outcomes did not improve as sharply in relative terms. This cohort remains vulnerable, particularly due to disrupted transitions related to education, employment, and social development. Longitudinal follow-up is needed to assess whether these early disruptions will lead to more persistent adverse outcomes in later adulthood.
Finally, the relatively higher levels of anxiety and depression in youth in the UK during the pandemic has been attributed to disruptions in the routines of life \citep{xiong2020impact,pierce2021mental}.  In particular changes in socialization, social isolation, remote learning (schools, universities) and reduced opportunities to work has played a part.
Despite these abovemetioned inequalities the literature does point to the fact the both the elderly and youth experienced poor mental health during the pandemic \cite{webb2020COVID}.

\subsection{Sex}
Women tended to have poorer mental health relative to men during the pandemic, and their mental health worsened more than men in relation to their pre-pandemic baseline. 
These disparities may be influenced by the increased caregiving responsibilities, employment insecurity, and exposure to domestic violence that disproportionately affected women during the pandemic. For instance, women were more likely to be employed in sectors shut down by lockdowns (e.g., hospitality, care work) and more likely to carry the burden of home-schooling and unpaid care work (ONS, 2021). Such role strain may have compounded existing gender inequalities and contributed to the observed mental health differences.

Intersectionalities regarding youth and sex are also potentially in play given the fact that our findings indicate that women, and separately, youth exhibit relatively poorer mental health compared to their respective reference categories during the pandemic.
Other researchers have indicated that the impact of {\covid} on youth has not been gender neutral \citep{hoyt2023internalizing}, and policies which are informed by existing intersectionalities within society are needed. Various studies \citep{oyafunke2021gender, parry2021shadow}  are highlighting the issue of gendered impact of \covid. {\covid} is said to have a “devastating” effect on women \citep{cousins2020COVID} and our findings suggest that they experienced worse mental health during this pandemic.  
Researchers point to the disproportionate shouldering of caregiving and household tasks \citep{liu2020prevalence, pieh2020effect, hiekel2022mental} as a possible reasons for the heavier impact of the pandemic on women.



\subsection{Ethnicity}
We did not find any statistically significant associations with mental health during the pandemic, and although both groups witness worsensed mental health from their respective pre-pandemic baseline, we did not find any statistically significant difference in their change from baseline. Many researchers indicate that ethnic minorities fared worse in terms of  mental health during the pandemic \citep{van2022mental, bhugra2021COVID, smith2020COVID}. These are intertwined with having a high-risk job, income level, deprivation, crowded living space, and systemic racism.
The lack of significance can also be due to the relatively smaller numbers of non-white ethnicities in the data. 

\subsection{Country}
On average we noted that during the pandemic, individuals from Wales experienced poorer mental than those in England, and that individuals from Northern Ireland experienced better mental health than England. Additionally, we note that mental health of individuals in Northern Ireland did not change as much as the other countries when compared to their respective pre-pandemic baslines.   
Possible reasons for these observations could be that the participants were based in different regions of deprivation or the areas where they were based were associated with varying levels of {\covid} restrictions.  Other possible reasons could be the fact that the lockdown rules differed at times between Northern Ireland and England, especially when the tiered restrictions were applied in England \citep{smith2022tiered} . 
Overall \cite{hubbard2021rurality}point out that in their research, living in an urban area or deprived area, or not having much access to outside space/green space(in the pandemic) was linked to worse psychological distress.

\subsection{Employment status}
Individuals who were unemployed had worse mental health during the pandemic than those who were employed.   However, when comparing their change in mental health relation to their respective baseline, employed individuals were affected more than unemployed individuals. 
In the literature, some researchers indicate that loss of employment during the pandemic led to poorer mental health \citep{lee2021estimating}. Unemployment \citep{ruffolo2021employment} and being on furlough \citep{davillas2021first} have been implicated in contributing to mental stress during the pandemic.
There is a lot of scope for investigating this area of focus especially within the context of furloughing and working reduced hours during the pandemic. However, we are mindful that the link between employment status on mental health \citep{murphy1999effect} already existed prior to the {\covid} pandemic.

\subsection{Income quintiles}
Individuals with an income greater than those in the poorest income quintile (Q1)  were found to have better mental health than those with a total household income at Q1 or lower.  Possible reasons for this could be due to not having access to paid leave or mental health services \citep{singh2020impact,xiong2020impact} and that that those in lower quintile were less likely to be able to work remotely \citep{bonacini2021working}.   Recall  that the income variable was not used in the {\GAMdiff} model. 

\subsection{Individuals not living with a partner}
The  individuals not living with a partner on average had worse mental health compared to those who lived with a partner, and their mental health worsened . more in relation to the respective pre-pandemic baselines.
Various researchers indicate that the presence of a partner or family members in the household contributed to better mental health \citep{sisson2022benefits} potentially because of the social interaction (no need to isolate) and opportunities to share.  These relationships were described as being “protective”.
When comparing mental health scores between baseline and {\covid} waves, the difference was found to be statistically significantly greater for individuals who did not live with a partner.  From our exploratory analyses (see Figure \ref{trend:figs}), we know that the mental health scores for those living with a partner and those not living with a partner increased (relative to the baseline) during the pandemic, so we can assume that those not living with a partner experienced the largest increase in mental health scores (translating to the largest worsening in mental health) during that time.  

The variable which we used specifies living with a partner and we acknowledge that there could be individuals living in a large household whilst not living with a partner and those who live alone.

\subsection{Living with children under age 16}
We did not find any statistically significant difference in mental health for individuals living with children under 16 compared to those who are not, but when considering the difference in mental health in relation to their respective pre-pandemic baseline, we observed that mental health worsened  (relative to those not living with children under 16) for individuals living with children under 16 and then worsened further if living with 2 or more children under 16. 
A similar effect was observed for the investigation of the impact of this variable on mental health during the pandemic, despite not being statistically significant. Potential reasons for these abovementioned findings include the responsibilities involved in caring for young children during the pandemic coupled with the fact that some adults needed to provide home-schooling during the pandemic \citep{zhang2022household}.
Notably, some researchers reported that there were individuals living with children under 16 who reported lower stress (better mental health) during the pandemic \citep{smail2020associations} because they received support for caring for their children.

\subsection{People with long-term illnesses}
During the pandemic, people with long-term illnesses were found to have poorer mental health compared to those with a long-term illness, and also their mental health worsened more in relation to their respective pre-pandemic baseline.  
Various researchers confirm this observation and point to possible reasons such as access to regular healthcare \citep{shevlin2020anxiety} and fear of contracting {\covid}  \citep{iob2020levels}.

\section{Conclusion}
\paragraph{Summary}
Our study has identified various mental health vulnerabilities in the UK population during the {\covid} pandemic. This suggests pre-existing inequalities within and between different demographic groupings within society. As a result of these, public health policies aimed at improving overall mental health would need to take these inequalities into consideration.

\paragraph{Strengths}
Generalized Additive Models allow for the investigation of the non-linear trend over time and various sociodemographic factors influencing mental health. Furthermore, we explore the change in mental health between the baseline period and the pandemic period and the association with a series of sociodemographic dimensions. The use of the UKHLS’s {\covid} waves allows accessing data that is collected more regularly during the (for 2020 and 2021).

\paragraph{Limitations}
One limitation of the analysis is that since the trajectories for the individuals are very different, a single general trend might not provide a good fit for every individual, and for each variable, considering a fixed effect throughout all waves can be a strong assumption. 
Another limitation is that the study does not consider a wide temporal trend which precedes the onset of the {\covid} pandemic. Finally, various studies [Wirz-Justice, 2018] indicate the importance of seasonality on mood and mental health. Our study assesses the trends observed during the pandemic and inequalities therein. As such, seasonality effects are not considered since the study period does not include repeated seasonal timepoints. 

\paragraph{Policy implications}
Mental health and wellbeing is a crucial  component of health with significant consequences for the social and economic wellbeing of individuals. This study shows the deleterious effects of {\covid} on the mental wellbeing of the population. It finds that the impact of the pandemic varied between social groups, with women, youth, Wales, unemployed, those with relatively low household income, those not living with a partner, those living with children under 16 years of age and those with long-term illness bearing the brunt. Young, unemployed and single individuals along with women would be of key interest to policy makers, since these represent concentrated demographic groups that deserve attention. Furthermore, intersectionality should be incorporated into the development of mental health interventions, since not only do women experience relatively poorer mental health than men, but also do parents with children under 16 years of age and single individuals.

\subsection{Declarations}

\begin{itemize}
\item	Ethics approval and consent to participate.
We have obtained ethics approval from the University of Edinburgh (Health in Social Science – Nursing Studies committee).  Our ethics application reference is $NUST016s$
\item	Consent for publication: Not applicable
\item	Availability of data and materials: The dataset analysed during the current study is publicly available via the UK Data Service.  The UKHLS (\url{https://www.understandingsociety.ac.uk/}) dataset which is a longitudinal dataset.
\item	Competing interests: The authors declare that they have no competing interests.
\item	Funding: College of Arts, Humanities and Social Sciences (CAHSS) Challenge Investment Fund (2021). Project Title: "Investigating mental wellbeing along the lifecourse (1991-2020) and during the {\covid} pandemic.", University of Edinburgh, UK. 
\item	Authors' contributions:
All authors read and approved the final manuscript.
\item	Acknowledgements
N/A
\item	Authors' contribution:
GN: conceptualization, funding acquisition, drafting, design of work.
ER: analysis, interpretation of data, creating of R code, drafting.
VP: conceptualization, funding acquisition, analysis, interpretation of data, creating of R code, drafting.
KM: analysis, interpretation of data, creating of R code.
CC: conceptualization, funding acquisition, drafting, design of work.
BL: conceptualization, funding acquisition, drafting, design of work.
AM: conceptualization, funding acquisition, drafting, design of work.
SS: conceptualization, funding acquisition, drafting, interpretation of data, design of work
SW:analysis, creating of R code.
\end{itemize}

\bibliographystyle{plainnat}
\bibliography{ukhlsrefs}

@article{bonacini2021working,
  title={Working from home and income inequality: risks of a ‘new normal’with COVID-19},
  author={Bonacini, Luca and Gallo, Giovanni and Scicchitano, Sergio},
  journal={Journal of population economics},
  volume={34},
  number={1},
  pages={303--360},
  year={2021},
  publisher={Springer}
}

@article{singh2020impact,
  title={Impact of COVID-19 and lockdown on mental health of children and adolescents: A narrative review with recommendations},
  author={Singh, Shweta and Roy, Deblina and Sinha, Krittika and Parveen, Sheeba and Sharma, Ginni and Joshi, Gunjan},
  journal={Psychiatry research},
  volume={293},
  pages={113429},
  year={2020},
  publisher={Elsevier}
}

@article{van2022mental,
  title={The mental health experiences of ethnic minorities in the UK during the Coronavirus pandemic: A qualitative exploration},
  author={Van Bortel, Tine and Lombardo, Chiara and Guo, Lijia and Solomon, Susan and Martin, Steven and Hughes, Kate and Weeks, Lauren and Crepaz-Keay, David and McDaid, Shari and Chantler, Oliver and others},
  journal={Frontiers in public health},
  volume={10},
  pages={875198},
  year={2022},
  publisher={Frontiers Media SA}
}

@article{bhugra2021covid,
  title={COVID-19 pandemic, mental health care, and the UK},
  author={Bhugra, Dinesh and Molodynski, Andrew and Gnanapragasam, Sam Nishanth},
  journal={Industrial psychiatry journal},
  volume={30},
  number={Suppl 1},
  pages={S5--S9},
  year={2021},
  publisher={Medknow}
}

@article{smith2022tiered,
  title={Tiered restrictions for COVID-19 in England: knowledge, motivation and self-reported behaviour},
  author={Smith, LE and Potts, HWW and Amlot, R and Fear, NT and Michie, Susan and Rubin, GJ},
  journal={Public Health},
  volume={204},
  pages={33--39},
  year={2022},
  publisher={Elsevier}
}

@article{hubbard2021rurality,
  title={Are rurality, area deprivation, access to outside space, and green space associated with mental health during the COVID-19 pandemic? A cross sectional study (CHARIS-E)},
  author={Hubbard, Gill and Daas, Chantal den and Johnston, Marie and Murchie, Peter and Thompson, Catharine Ward and Dixon, Diane},
  journal={International Journal of Environmental Research and Public Health},
  volume={18},
  number={8},
  pages={3869},
  year={2021},
  publisher={MDPI}
}

@misc{smith2020covid,
  title={COVID-19, mental health and ethnic minorities},
  author={Smith, Katharine and Bhui, Kamaldeep and Cipriani, Andrea},
  journal={BMJ Ment Health},
  volume={23},
  number={3},
  pages={89--90},
  year={2020},
  publisher={Royal College of Psychiatrists}
}

@article{iob2020levels,
  title={Levels of severity of depressive symptoms among at-risk groups in the UK during the COVID-19 pandemic},
  author={Iob, Eleonora and Frank, Philipp and Steptoe, Andrew and Fancourt, Daisy},
  journal={JAMA network open},
  volume={3},
  number={10},
  pages={e2026064--e2026064},
  year={2020},
  publisher={American Medical Association}
}

@article{shevlin2020anxiety,
  title={Anxiety, depression, traumatic stress and COVID-19-related anxiety in the UK general population during the COVID-19 pandemic},
  author={Shevlin, Mark and McBride, Orla and Murphy, Jamie and Miller, Jilly Gibson and Hartman, Todd K and Levita, Liat and Mason, Liam and Martinez, Anton P and McKay, Ryan and Stocks, Thomas VA and others},
  journal={BJPsych open},
  volume={6},
  number={6},
  pages={e125},
  year={2020},
  publisher={Cambridge University Press}
}

@article{sisson2022benefits,
  title={The benefits of living with close others: A longitudinal examination of mental health before and during a global stressor},
  author={Sisson, Natalie M and Willroth, Emily C and Le, Bonnie M and Ford, Brett Q},
  journal={Clinical Psychological Science},
  volume={10},
  number={6},
  pages={1083--1097},
  year={2022},
  publisher={Sage Publications Sage CA: Los Angeles, CA}
}

@article{lee2021estimating,
  title={Estimating influences of unemployment and underemployment on mental health during the COVID-19 pandemic: who suffers the most?},
  author={Lee, Jungeun Olivia and Kapteyn, Arie and Clomax, Adriane and Jin, Haomiao},
  journal={Public Health},
  volume={201},
  pages={48--54},
  year={2021},
  publisher={Elsevier}
}

@article{ruffolo2021employment,
  title={Employment uncertainty and mental health during the COVID-19 pandemic initial social distancing implementation: a cross-national study},
  author={Ruffolo, Mary and Price, Daicia and Schoultz, Mariyana and Leung, Janni and Bonsaksen, Tore and Thygesen, Hilde and Geirdal, Amy {\O}stertun},
  journal={Global Social Welfare},
  volume={8},
  pages={141--150},
  year={2021},
  publisher={Springer}
}

@article{davillas2021first,
  title={The first wave of the COVID-19 pandemic and its impact on socioeconomic inequality in psychological distress in the UK},
  author={Davillas, Apostolos and Jones, Andrew M},
  journal={Health economics},
  volume={30},
  number={7},
  pages={1668--1683},
  year={2021},
  publisher={Wiley Online Library}
}

@article{zhang2022household,
  title={Household chaos and caregivers’ and young children’s mental health during the COVID-19 pandemic: A mediation model},
  author={Zhang, Xiao},
  journal={Journal of child and family studies},
  volume={31},
  number={6},
  pages={1547--1557},
  year={2022},
  publisher={Springer}
}

@article{smail2020associations,
  title={Associations of household structure and presence of children in the household with mental distress during the COVID-19 pandemic},
  author={Smail, Emily and Riehm, Kira and Veldhuis, Cindy and Johnson, Renee and Holingue, Calliope and Stuart, Elizabeth A and Kalb, Luke and Thrul, Johannes},
  year={2020},
  publisher={OSF},
journal={OSF}
}

@article{hiekel2022mental,
  title={Mental health before and during the COVID-19 pandemic: The role of partnership and parenthood status in growing disparities between types of families},
  author={Hiekel, Nicole and K{\"u}hn, Mine},
  journal={Journal of Health and Social Behavior},
  volume={63},
  number={4},
  pages={594--609},
  year={2022},
  publisher={Sage Publications Sage CA: Los Angeles, CA}
}

@article{pieh2020effect,
  title={The effect of age, gender, income, work, and physical activity on mental health during coronavirus disease (COVID-19) lockdown in Austria},
  author={Pieh, Christoph and Budimir, Sanja and Probst, Thomas},
  journal={Journal of psychosomatic research},
  volume={136},
  pages={110186},
  year={2020},
  publisher={Elsevier}
}

@article{liu2020prevalence,
  title={Prevalence and predictors of PTSS during COVID-19 outbreak in China hardest-hit areas: Gender differences matter},
  author={Liu, Nianqi and Zhang, Fan and Wei, Cun and Jia, Yanpu and Shang, Zhilei and Sun, Luna and Wu, Lili and Sun, Zhuoer and Zhou, Yaoguang and Wang, Yan and others},
  journal={Psychiatry research},
  volume={287},
  pages={112921},
  year={2020},
  publisher={Elsevier}
}

@article{webb2020covid,
  title={Covid-19 lockdown: a perfect storm for older people’s mental health},
  author={Webb, Lucy},
  journal={Journal of psychiatric and mental health nursing},
  volume={28},
  number={2},
  pages={300},
  year={2020}
}

@article{xiong2020impact,
  title={Impact of COVID-19 pandemic on mental health in the general population: A systematic review},
  author={Xiong, Jiaqi and Lipsitz, Orly and Nasri, Flora and Lui, Leanna MW and Gill, Hartej and Phan, Lee and Chen-Li, David and Iacobucci, Michelle and Ho, Roger and Majeed, Amna and others},
  journal={Journal of affective disorders},
  volume={277},
  pages={55--64},
  year={2020},
  publisher={Elsevier}
}

@article{coco2023psychosocial,
  title={Psychosocial predictors of trajectories of mental health distress during the COVID-19 pandemic: A four-wave panel study},
  author={Coco, Gianluca Lo and Salerno, Laura and Albano, Gaia and Pazzagli, Chiara and Lagetto, Gloria and Mancinelli, Elisa and Freda, Maria Francesca and Bassi, Giulia and Giordano, Cecilia and Gullo, Salvatore and others},
  journal={Psychiatry Research},
  volume={326},
  pages={115262},
  year={2023},
  publisher={Elsevier}
}

@article{wels2022mental,
  title={Mental and social wellbeing and the UK coronavirus job retention scheme: Evidence from nine longitudinal studies},
  author={Wels, Jacques and Booth, Charlotte and Wielgoszewska, Bo{\.z}ena and Green, Michael J and Di Gessa, Giorgio and Huggins, Charlotte F and Griffith, Gareth J and Kwong, Alex SF and Bowyer, Ruth CE and Maddock, Jane and others},
  journal={Social Science \& Medicine},
  volume={308},
  pages={115226},
  year={2022},
  publisher={Elsevier}
}

@article{wang2024impact,
  title={The impact of reduced working hours and furlough policies on workers’ mental health at the onset of COVID-19 pandemic: a longitudinal study},
  author={Wang, Senhu and Kamer{\=a}de, Daiga and Bessa, Ioulia and Burchell, Brendan and Gifford, Jonny and Green, Melanie and Rubery, Jill},
  journal={Journal of Social Policy},
  volume={53},
  number={3},
  pages={702--726},
  year={2024},
  publisher={Cambridge University Press}
}

@techreport{smyth2022disrupted,
  title={Disrupted transitions? Young adults and the COVID-19 pandemic},
  author={Smyth, Emer and Nolan, Anne},
  year={2022},
  institution={Research Series}
}

@inproceedings{colombo2024effect,
  title={The Effect of COVID-19 on Middle-Aged Adults’ Mental Health: A Mixed-Method Case--Control Study on the Moderating Effect of Cognitive Reserve},
  author={Colombo, Barbara and Fusi, Giulia and Christopher, Kenneth B},
  booktitle={Healthcare},
  volume={12},
  pages={163},
  year={2024},
  organization={MDPI}
}

@article{hoyt2023internalizing,
  title={Internalizing the COVID-19 pandemic: Gendered differences in youth mental health},
  author={Hoyt, Lindsay Till and Dotson, Miranda P and Suleiman, Ahna Ballonoff and Burke, Natasha L and Johnson, Jasmine B and Cohen, Alison K},
  journal={Current Opinion in Psychology},
  volume={52},
  pages={101636},
  year={2023},
  publisher={Elsevier}
}

@article{oyafunke2021gender,
  title={Gender-based violence and Covid-19: The shadow pandemic in Africa},
  author={Oyafunke-Omoniyi, Comfort O and Adisa, Isaiah and Obileye, Abolaji A},
  journal={Gendered Perspectives on Covid-19 Recovery in Africa: Towards Sustainable Development},
  pages={55--71},
  year={2021},
  publisher={Springer}
}

@article{parry2021shadow,
  title={The shadow pandemic: Inequitable gendered impacts of COVID-19 in South Africa},
  author={Parry, Bianca Rochelle and Gordon, Errolyn},
  journal={Gender, Work \& Organization},
  volume={28},
  number={2},
  pages={795--806},
  year={2021},
  publisher={Wiley Online Library}
}

@article{murphy1999effect,
  title={The effect of unemployment on mental health},
  author={Murphy, Gregory C and Athanasou, James A},
  journal={Journal of Occupational and organizational Psychology},
  volume={72},
  number={1},
  pages={83--99},
  year={1999},
  publisher={Wiley Online Library}
}

@article{perelli2023better,
  title={For better or worse: Economic strain, furlough, and relationship quality during the Covid-19 lockdown},
  author={Perelli-Harris, Brienna and Chao, Shih-Yi and Berrington, Ann},
  journal={Journal of Marriage and Family},
  year={2023},
  publisher={Wiley Online Library}
}

@article{perelli2020couples,
  title={Couples in crisis: how the government's furlough scheme has protected relationships during the Covid-19 pandemic},
  author={Perelli-Harris, Brienna and Chao, Shih-Yi and Berrington, Ann},
  year={2020},
  publisher={ESRC Centre for Population Change},
journal={ESRC}
}

@article{daly2022longitudinal,
  title={Longitudinal changes in mental health and the COVID-19 pandemic: Evidence from the UK Household Longitudinal Study},
  author={Daly, Michael and Sutin, Angelina R and Robinson, Eric},
  journal={Psychological medicine},
  volume={52},
  number={13},
  pages={2549--2558},
  year={2022},
  publisher={Cambridge University Press}
}

@article{pierce2021mental,
  title={Mental health responses to the COVID-19 pandemic: a latent class trajectory analysis using longitudinal UK data},
  author={Pierce, Matthias and McManus, Sally and Hope, Holly and Hotopf, Matthew and Ford, Tamsin and Hatch, Stephani L and John, Ann and Kontopantelis, Evangelos and Webb, Roger T and Wessely, Simon and others},
  journal={The Lancet Psychiatry},
  volume={8},
  number={7},
  pages={610--619},
  year={2021},
  publisher={Elsevier}
}

@article{mollica2004mental,
  title={Mental health in complex emergencies},
  author={Mollica, Richard F and Cardozo, B Lopes and Osofsky, Howard J and Raphael, Beverley and Ager, Alastair and Salama, Peter},
  journal={The Lancet},
  volume={364},
  number={9450},
  pages={2058--2067},
  year={2004},
  publisher={Elsevier}
}

@article{jackson2017hair,
  title={Hair cortisol and adiposity in a population-based sample of 2,527 men and women aged 54 to 87 years},
  author={Jackson, Sarah E and Kirschbaum, Clemens and Steptoe, Andrew},
  journal={Obesity},
  volume={25},
  number={3},
  pages={539--544},
  year={2017},
  publisher={Wiley Online Library}
}

@article{silove2013adapt,
  title={The ADAPT model: a conceptual framework for mental health and psychosocial programming in post conflict settings},
  author={Silove, Derrick and others},
  journal={Intervention},
  volume={11},
  number={3},
  pages={237--248},
  year={2013},
  publisher={Medknow Publications}
}

@article{rahman2021suicidal,
  title={Suicidal behaviors and suicide risk among Bangladeshi people during the COVID-19 pandemic: an online cross-sectional survey},
  author={Rahman, Md Estiar and Al Zubayer, Abdullah and Bhuiyan, Md Rifat Al Mazid and Jobe, Mary C and Khan, Md Kamrul Ahsan},
  journal={Heliyon},
  volume={7},
  number={2},
  pages={e05937},
  year={2021},
  publisher={Elsevier}
}

@article{davillas,
  title={The first wave of the COVID-19 pandemic and its impact on socioeconomic inequality in psychological distress in the UK},
  author={Davillas, Apostolos and Jones, Andrew M},
  journal={Health economics},
  volume={30},
  number={7},
  pages={1668--1683},
  year={2021},
  publisher={Wiley Online Library}
}

@article{taylor2006hypochondriasis,
  title={Hypochondriasis},
  author={Taylor, Steven and Asmundson, Gordon JG},
  journal={Practitioner’s Guide to Evidence-Based Psychotherapy},
  pages={313--323},
  year={2006},
  publisher={Springer}
}

@article{makhanova2020behavioral,
  title={Behavioral immune system linked to responses to the threat of COVID-19},
  author={Makhanova, Anastasia and Shepherd, Melissa A},
  journal={Personality and Individual Differences},
  volume={167},
  pages={110221},
  year={2020},
  publisher={Elsevier}
}

@article{cousins2020covid,
  title={COVID-19 has “devastating” effect on women and girls},
  author={Cousins, Sophie},
  journal={The Lancet},
  volume={396},
  number={10247},
  pages={301--302},
  year={2020},
  publisher={Elsevier}
}

@article{barford2021youth,
  title={Youth employment in times of COVID: A global review of COVID-19 policy responses to tackle (un) employment and disadvantage among young people. ILO: Geneva.},
  author={Barford, Anna and Coutts, Adam and Sahai, Garima},
  year={2021},
  publisher={International Labour Organization},
journal={ILO}
}
\end{document}